\documentclass{jfm}

\usepackage{graphicx}
\usepackage{newtxtext}
\usepackage{newtxmath}
\usepackage{natbib}
\usepackage{hyperref}
\hypersetup{
    colorlinks = true,
    urlcolor   = blue,
    citecolor  = blue,
}

\newcommand{\RomanNumeralCaps}[1]

\usepackage{graphicx}
\usepackage[dvipsnames]{xcolor}
\usepackage{amsmath}
\usepackage{amssymb}

\shorttitle{Convection in porous--fluid layers}
\shortauthor{T. Le Reun and D.R Hewitt}

\title{High-Rayleigh-number convection in porous--fluid layers 
}

\author{Thomas Le Reun \aff{1}
  \corresp{\email{tl402@cam.ac.uk}} \and
 Duncan R. Hewitt \aff{2}
 }

\affiliation{\aff{1} DAMTP, University of Cambridge, Wilberforce Road, Cambridge CB3 0WA, UK
\aff{2} Department of Mathematics, University College London, UK 
}

\xdefinecolor{RED}{named}{black}

\newcommand{\mbf}{\boldsymbol}

\newcommand{\ie}{\textit{i.e.}~}
\newcommand{\bu}{\mbf{u}}
\newcommand{\bU}{\mbf{U}}

\newcommand{\taud}{\tau_{\mathrm{diff}}}

\newcommand{\Ra}{\mathrm{Ra}}

\newcommand{\Da}{\mathrm{Da}}

\newcommand{\Pei}{\mathrm{Ra}^{-1/2}}

\newcommand{\Nu}{\mathrm{Nu}}
\newcommand{\Cf}{\mathcal{C}_f}
\newcommand{\Cp}{\mathcal{C}_p}
\newcommand{\N}{\mathrm{N}}
\newcommand{\R}{\mathrm{R}}

\newcommand{\RB}{Rayleigh-B\'enard}
\newcommand{\hf}{\hat{h}_f}
\newcommand{\hh}{\hat{h}}
\newcommand{\hp}{\hat{h}_p}

\newcommand{\hR}{\hat{\mathrm{R}}}
\newcommand{\Ja}{\left\langle J \right\rangle}
\newcommand{\Jav}{\left\langle J \right\rangle}
\newcommand{\wrms}{w_{\mathrm{rms}}}

\newcommand{\RaDa}{\ensuremath{\sqrt{\mathrm{Ra}} \Da}}

\usepackage{graphicx}

\graphicspath{{./}}

\usepackage{cleveref}
\crefname{equation}{}{}
\crefname{figure}{figure}{figures}
\begin{document}
\renewcommand{\ref}{\cref}

\maketitle

\begin{abstract}

We present a numerical study of convection in a horizontal layer comprising a fluid-saturated porous bed overlain by an unconfined fluid layer.
Convection is driven by a vertical, destabilising temperature difference applied across the whole system, as in the canonical Rayleigh--B\'enard problem. 
Numerical simulations are carried out using a single-domain formulation of the two-layer problem based on the Darcy--Brinkman equations.
We explore the dynamics and heat flux through the system in the limit of large Rayleigh number, but small Darcy number, such that the flow exhibits vigorous convection in both the porous and the unconfined fluid regions, while the porous flow still remains strongly confined and governed by Darcy's law. 
We demonstrate that the heat flux and average thermal structure of the system can be predicted using previous results of convection in individual fluid or porous layers. 
We revisit a controversy about the role of subcritical ``penetrative convection'' in the porous medium, and confirm that such induced flow does not contribute to the heat flux through the system. 
Lastly, we briefly study the temporal coupling between the  two layers and find that the turbulent fluid convection above acts as a low-pass filter on the longer-timescale variability of convection in the porous layer. 

\end{abstract}

\section{Introduction}

Heat transfer driven by flow exchange between a fluid-saturated porous bed and an overlying unconfined fluid arises in a variety of systems in engineering and geophysics. This is the case, for example, in various industrial cooling systems found in nuclear power generation, microelectronics or chemical engineering that require the circulation of fluid from an open channel into a fragmented medium \citep{dhueppe_coupling_2012,chandesris_direct_2013,su_macroscopic_2015}.
A similar situation occurs during the progressive solidification of multi-component fluids, which creates a mushy solid through which liquid flows to transport heat and solute \citep{worster_convection_1997}. In geophysical contexts, this phenomenon is encountered below sea ice \citep{wells_mushy-layer_2019} and in the Earth's core \citep{huguet_structure_2016}.
This work is particularly motivated by the physics of hydrothermal circulation, where a water-saturated, porous bed that is heated from below exhibits thermal convection that, in turn drives buoyant plumes and convective motion in the overlying ocean. 
As well as being a well-known feature of the Earth's ocean, there is evidence of on-going hydrothermal activity under the ice crust of icy satellites such as Jupiter's moon Europa \citep{goodman_hydrothermal_2004} or Saturn's moon Enceladus \citep{hsu_ongoing_2015}.
Unlike on Earth, the entire heat budget of these bodies is believed to be controlled by hydrothermal convection and, in particular, by the manner in which heat is transported through the rocky core and into the overlying oceans beneath their icy crusts \citep{travis_whole-moon_2012,travis_keeping_2015,nimmo_geophysics_2018}.
Most previous works in this area have focussed on either the flow in the porous medium alone or on that in unconfined fluid alone, with the coupling between them modelled by a parametrised boundary condition. 
This is particularly the case for hydrothermal activity, for which there are numerous studies focussed either on the structure of the flow in the porous layer (see for instance \cite{fontaine_two-dimensional_2007,coumou_structure_2008,choblet_powering_2017,le_reun_internally_2020}, among others), or on the buoyant plumes created in the ocean \citep{goodman_hydrothermal_2004,woods_turbulent_2010,goodman_numerical_2012}, or on the induced large-scale oceanic circulation \citep{soderlund_ocean-driven_2014,
soderlund_ocean_2019,amit_cooling_2020}.   
\cite{travis_whole-moon_2012} included both layers, but resorted to an enhanced diffusivity to parametrise flows in the sub-surface ocean to make calculations tractable. 
In all these cases, questions remain about how reasonable it is to use these parameterised boundary conditions rather than resolve both layers, and about how the dynamics of flow in each layer communicate and influence the flow in the other layer. Addressing these questions is the focus of this paper.

Works that explicitly focus on the coupled transport across a porous and fluid layer are more numerous in engineering settings. However, they tend to either be focused on situations where inertial effects in the interstices of the porous layer play an important role \citep{dhueppe_boundary_2011,chandesris_direct_2013} or on regimes where heat is mainly transported by diffusion through the porous layer \citep{poulikakos_high_1986,
chen_onset_1988,
chen_convection_1992,
bagchi_natural_2014,
su_macroscopic_2015}, 
\textcolor{RED}{or where restricted to the onset of convective instabilities
\citep{chen_onset_1988,
hirata_linear_2007,hirata_stability_2009a,hirata_stability_2009}.
}
  In general, these studies are difficult to apply to geophysical contexts, and particularly to hydrothermal circulation: here, the large spatial and temperature scales and the typically relatively low permeabilities are such that the porous region can be unstable to strong convection while the flow through it remains inertia-free and well described by Darcy's law.  In such a situation, there can be vastly different timescales of motion between the unconfined fluid, which exhibits rapid turbulent convection, and the porous medium, where the convective flow through the narrow pores is much slower. This discrepancy in timescales presents a challenge for numerical modelling, which is perhaps why this limit has not been explored until now. 


%

In this paper, we explore thermal convection in a two-layer system comprising a porous bed overlain by an unconfined fluid. In particular, we focus on situations in which the driving density difference, as described by the dimensionless Rayleigh number, is large and heat is transported through both layers by convective flow, although for completeness we include cases in which there is no convection in the porous layer. The permeability of the porous medium, as described by the dimensionless Darcy number, is small enough that the flow through the medium is always inertia free and controlled by Darcy's law. 
%
As in some previous studies of this setup \citep{poulikakos_high_1986,chen_onset_1988,bagchi_natural_2014}, we consider the simplest idealised system in which natural convection occurs, that is, a two-layer Rayleigh-B\'enard cell. In this setup, the base of the porous medium is heated and the upper surface of the unconfined fluid layer is cooled to provide a fixed destabilising density difference across the domain. Flow in such a system attains a statistically steady state, which allows for investigation of the fluxes, temperature profiles and dynamics of the flow in each layer, and of the coupling between the layers. 

We carry out numerical simulations of this problem in two-dimensions using a single-domain formulation of the two-layer problem based on the Darcy-Brinkman equations \citep{le_bars_interfacial_2006}.
Using efficient pseudo-spectral methods, we are able to reach regimes where thermal instabilities are fully developed in both the porous and the fluid layer. 
We demonstrate how to use previous results on thermal convection in individual fluid or porous layers to infer predictions of the heat flux and the temperature at the interface between the layers in our system.
In addition, we revisit a long-standing controversy on the role of ``penetrative convection'', \ie flow in the porous medium that is actively driven by fluid convection above, and confirm that it is negligible in the limit where the pore scale is small compared to the size of the system. 
Lastly, we briefly address the temporal coupling between both layers and explore how fluid convection mediates the variability of porous convection. 
The paper is organised as follows.  
The setup and governing equations are introduced in \ref{section1}, where we also outline the main approximations of our model and, importantly, the limits on its validity. 
After presenting the general behaviour of the two-layer system and how it changes when the porous layer becomes unstable to convection (\ref{sec:overview}), we show how previous works on convection can be used to predict the thermal structure of the flow and the heat it transports (\ref{sec_theory}). In \ref{sec:temporal} we discuss the temporal variability of two-layer convection, before summarizing our findings and their geophysical implications in \ref{sec:conclusion}.

\section{Governing equations and numerical methods}
\label{section1}

\begin{figure}
\centerline{
\includegraphics[width=0.45\linewidth]{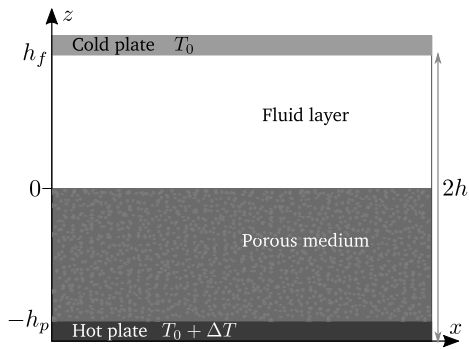}
}
\caption{Schematic cartoon of the set-up considered in this paper. In almost all cases considered here, we take the layer depths to be equal, so $h_p=h_f = h$. }
\label{fig:setup}
\end{figure}

\subsection{The Darcy--Brinkman model}
\label{sec:DarcyBrinkmanmodel}

Consider a two-dimensional system comprising a fluid-saturated porous medium of depth $h_p$ underlying an unconfined fluid region of depth $h_f$. 
We locate the centre of the cell at height $z=0$, such that the whole system lies between $-h_p \leqslant z  \leqslant h_f$, as depicted in \ref{fig:setup}, and we introduce the length scale $h = (h_p+h_f)/2$. For the majority of this paper we focus on the case of equal layer depths, where $h_p = h_f = h$. 
%

The fluid has a dynamic viscosity $\eta$ and density $\rho$, and the porous medium is characterised by a uniform permeability $K_0$ and porosity $\phi_0< 1$. 
We extend the definition of the porosity - that is, the local volume fraction of fluid - to the whole domain by introducing the step function
\begin{equation}
\phi(z) = \left\{\begin{array}{cc}
		\phi_0 & z<0 \\
		1 & z\geq 0 
		\end{array} \right.
		\label{phi_eq}
\end{equation}
Given local fluid velocity $\bU_{\ell}$, the flux $\bU = \phi \bU_{\ell}$ reduces to the usual Darcy flux in the porous medium, and to the fluid velocity in the unconfined fluid layer.

The flow is assumed to be incompressible everywhere, so
\begin{equation}
\label{eq:mass_conservation}
\bnabla \cdot \bU = 0.
\end{equation}
In the fluid later, the flow is governed by the Navier--Stokes equation, 
\begin{equation}
\label{eq:Navier_Stokes}
\rho \left[\displaystyle \p_t \bU + \bU \cdot \bnabla \bU  \right]  = -  \displaystyle   \bnabla P + \mu \bnabla^2 \bU +  \rho  \mbf{g}
\end{equation}
where $P$ is the pressure, while in the porous layer, the flux instead obeys Darcy's law, 
\begin{equation}
\label{eq:Darcy_dimensional}
\bU = \frac{K}{\mu}\left(- \displaystyle   \bnabla P +  \rho  \mbf{g} \right).
\end{equation}

We simulate the two-layer system using a one-domain approach in which both porous and unconfined fluid regions are described by a single Darcy-Brinkman equation \citep{le_bars_interfacial_2006},
\begin{equation}
\label{eq:momentum_conservation}
\rho \left[\displaystyle \p_t \bU + \bU \cdot \bnabla \frac{\bU}{\phi} \right]  = - \phi \displaystyle   \bnabla P + \mu \bnabla^2 \bU + \phi \rho  \mbf{g} - \frac{\mu}{K} \phi  \bU.
\end{equation}
where $1/K(z)$ is a step function that goes from $1/K_0$ for $z<0$ to zero for $z>0$. 
%
%
%
As shown by \cite{le_bars_interfacial_2006}, the Darcy-Brinkman formulation of the two-layer problem can be retrieved by carrying out a coarse-grained average of the flow over a few typical pore scales $\sqrt{K_0}$. 
As a consequence, any spatial variation must have a typical length larger than $\sqrt{K}$ for the model to remain valid. 
The Navier--Stokes equation and Darcy's law are retrieved from the Darcy-Brinkman equation \ref{eq:momentum_conservation} in the unconfined fluid and porous layers, respectively. 
In the fluid layer, $\phi = 1$ and $1/K = 0$, which trivially gives the Navier-Stokes equation, whereas in the bulk of the porous medium, the damping term $-\mu \phi \bU /K $ dominates over the inertial and viscous forces (provided $K_0$ is sufficiently small), leading to a balance between the damping, pressure and buoyancy terms that yields Darcy's law.  
Just below the interface, however, viscous effects become important in the porous layer as the flow matches to the unconfined region above. Acceleration remains negligible, and the equations reduces to
\begin{equation}
\label{eq:interface_equation}
- \phi \displaystyle   \bnabla P + \mu \bnabla^2 \bU + \phi \rho  \mbf{g} - \frac{\mu}{K} \phi  \bU \sim 0.
\end{equation}
Balancing the viscous resistance and Darcy drag indicates that local viscous forces play a role over a length $\ell_r = \sqrt{K_0}/\phi_0$ below the the interface---\ie a few times the pore scale. 
These forces regularise the velocity profile between the unconfined fluid and the porous medium \textcolor{RED}{through a boundary layer of typical length $\ell_r$.}
\textcolor{RED}{
To conclude, we choose to model the two-layer system with a one-domain formulation via the Darcy-Brinkman equation. 
Another classical alternative formulation of the problem is the one introduced after \cite{beavers_boundary_1967} where the fluid and the porous layer are treated separately and their coupling is accounted by a semi-empirical boundary condition linking vertical velocity gradients and the velocity difference between the fluid and porous layers.
These different models all feature the regularisation boundary layer at the fluid-porous interface over a length $\sim \sqrt{K_0}$, which is corroborated by many experiments, in particular those of \cite{beavers_boundary_1967}. 
Yet, there are known discrepancies between these two approaches that may not be restricted to the interface, as shown by \cite{le_bars_interfacial_2006}.
In the specific case of two-layer convection, \cite{hirata_linear_2007,hirata_stability_2009a} have demonstrated that the use of one-domain or two-domain formulation lead to sightly different thresholds for the development of convective instabilities in the two-layer system. 
As pointed out by \cite{nield_convection_2017}, there is still debate on which of these formulations is the most adequate to model flows in porous-fluid layers, definitive empirical evidence being still awaiting. 
Although we choose to use the one-domain formulation of fluid-porous convection, we will show numerical results that can be understood without relying on the details of the interface flow were models might be inaccurate. 
}

\subsection{Heat transport}
We use the Boussinesq approximation to account for the effect of temperature-dependent density in the momentum equations: variations in temperature affect the buoyancy force but do not affect the fluid volume via conservation of mass.
We further assume that any changes in viscosity, diffusivity or permeability associated with temperature variation are negligible. 
%
%
Although some of these assumptions may be questionable in complex geophysical settings, they are made here to allow a focus on the basic physics of these two-layer convecting systems.
%
%
%
%

%
In particular, we restrict our study to linear variations of density with temperature according to 
\begin{equation}
\label{eq:linear_density}
\rho = \rho_0 \left( 1 - \alpha (T-T_0) \right)
\end{equation}
with $T_0$ being a reference temperature.
The momentum equation under the Boussinesq approximation follows from substituting \ref{eq:linear_density} into the buoyancy term of \ref{eq:momentum_conservation}, while letting $\rho = \rho_0$ in the inertial terms.
The temperature evolves according to an energy transport equation \citep{nield_heat_2013}
\begin{equation}
\label{eq:energy_conservation}
\overline{\phi}\p_t T + \bU \cdot \bnabla T = \kappa \Lambda \nabla^2 T 
\end{equation}
\begin{equation}
\displaystyle \mbox{with}~~~ \overline{\phi} \equiv \frac{(1-\phi) c_m \rho_m + \phi c \rho}{\rho c }, ~~~ \kappa \equiv \frac{\lambda}{\rho c}~~~ \mbox{and} ~~~  \Lambda \equiv \phi  + ( 1-\phi)\frac{\lambda_m}{\lambda} 
\end{equation}
where $c$ and $c_m$ are the heat capacity per unit of mass of the fluid and the porous matrix respectively, $\rho_m$ is the density of the porous matrix, and $\lambda$ and $\lambda_m$ are the thermal conductivities of the fluid and the porous matrix respectively. 
Equation \cref{eq:energy_conservation} assumes local thermal equilibrium between the porous matrix and the fluid.
%

\subsection{Boundary conditions}
We consider a closed domain with imposed temperature on the upper and lower boundaries, as in a classical Rayleigh--Benard cell (\ref{fig:setup}). Specifically, for the temperature we set
\begin{equation}
T(z=h) = T_0,
\qquad
T(z=-h) = T_0 + \Delta T,
\end{equation}
where $\Delta T >0$ is a constant. The upper and lower boundaries are rigid and impermeable, so 
\begin{equation}
\label{eq:bcs}
U(z=\pm h) = V(z=\pm h) = 0.
\end{equation}
Note that Darcy's law would only permit one velocity boundary condition on the boundary of the porous region at $z=-h$, but the higher-order derivative in the viscous term in \ref{eq:momentum_conservation} allows for application of the no-slip condition in \ref{eq:bcs} as well. 
\textcolor{RED}{
This extra condition will induce a boundary layer of thickness $\sim \sqrt{K_0}/\phi_0$ at the base of the domain, just like the boundary-layer region at the interface (see \ref{eq:interface_equation}).
It is not clear whether such a basal boundary layer should arise in experimental situations.
Nevertheless, it plays no dynamical role here provided it is thinner than any dynamical lengthscale of the flow (and, in particular, thinner than the thermal boundary layer that can form at the base of the domain, as discussed in \cref{sec:limits}.)
}
%

The horizontal boundaries of the domain are assumed to be periodic with the width of the domain kept constant at $4h$.

%

%

\subsection{Dimensionless equations and control parameters}
\label{sec:dimensionless_equations}
%

%
%
%

%
In order to extract the dimensionless equations that govern the two-layer system we use a ``free-fall'' normalisation of \ref{eq:momentum_conservation} and \ref{eq:energy_conservation}, based on the idea that a balance between inertia and buoyancy governs the behaviour of the fluid layer. 
Such a balance yields the free-fall velocity in the unconfined layer,
\begin{equation}
{U^*}^2 =  \alpha \Delta T g h,
\end{equation}
and the associated free-fall timescale is $T^* = h/U^*$. 
Scaling lengths with $h$, flux with $U^*$, time with $T^*$, temperature with $\Delta T$ and pressure with $U^*$, we arrive at dimensionless equations
%
%
 \begin{subequations}
 \label{eq:dimensionless_equations}
\begin{align}
&\p_t \bu + \bu \cdot \bnabla \frac{ \bu}{\phi}   = - \phi \bnabla p + \sqrt{\frac{\Pr}{\Ra}} \bnabla^2 \bu + \phi \theta  \mbf{e}_z - \frac{ \chi(z)}{\Da} \sqrt{\frac{\Pr}{\Ra}}  \phi \bu \label{eq:dl_mc_RaPr} \\
& \overline{\phi} \p_t \theta + \bu \cdot \bnabla \theta = \frac{\Lambda}{\sqrt{\Ra \Pr}} \nabla^2 \theta \label{eq:dl_ec_RaPr}\\
& \bnabla \cdot \bu = 0, \label{eq:dl_ec}
\end{align}
 \end{subequations}
where $\bu$ is the dimensionless flux, $\theta = (T-T_0)/\Delta T$ is the dimensionless temperature and $\chi(z)$ is a step function that jumps from $1$ in $z<0$ to $0$ in $z>0$. 
In these equations, we have introduced three dimensionless numbers: 
\begin{equation}
\label{parameters}
\Da \equiv \frac{K_0}{h^2}, ~~~ \Pr \equiv \frac{\nu}{\kappa} ~~~\mbox{and} ~~~ \Ra \equiv \frac{\alpha  g\Delta T h^3 }{\nu \kappa}~. 
\end{equation} 
The Darcy number $\Da$ is a dimensionless measure of the pore scale $\sqrt{K_0}$ relative to the domain size $h$, and is thus typically extremely small. 
The Rayleigh number quantifies the importance of the buoyancy forces relative to the viscous resistance in the unconfined fluid layer, and the focus of this work is on cases where $Ra \gg 1$. 
The Prandtl number compares momentum and heat diffusivities. The dimensionless layer depths $\hat{h}_p$ and $\hat{h}_f$ are also, in general, variables; as noted above, in the majority of computations shown here we set these to be equal so that $\hat{h}_p = \hat{h}_f = 1$, but we include the general case in the theoretical discussion in \ref{sec_theory}. 
%
%
%

Note that with this choice of scalings, the dimensionless velocity scale in the fluid layer is $O(1)$, compared with $O(\sqrt{\Ra} \Da /\sqrt{\Pr})$ in the porous layer.  Given these scales, we can also introduce a porous Rayleigh number $\Ra_p$ to describe the flow in the porous layer.  The porous Rayleigh number is the ratio between the advective and diffusive timescales in the porous medium, which, from the advection--diffusion ratio in \ref{eq:dl_ec_RaPr}, gives 
\begin{equation}
\label{Rap}
\Ra_p = \frac{\Ra \Da}{ \Lambda} = \frac{\alpha  g\Delta T K_0 h }{\nu \Lambda \kappa}~. 
\end{equation}

\subsection{Further simplifying assumptions}
We simplify the complexity of \ref{eq:dimensionless_equations} by noting that in the bulk of either the fluid or the porous regions, $\phi$ cancels out of the equations (see for instance \ref{eq:Navier_Stokes} and \ref{eq:Darcy_dimensional}). 
The porosity only affects \ref{eq:dimensionless_equations} in the narrow boundary-layer region immediately below the interface and at the base of the domain, where it controls the regularisation length $\sqrt{\Da}/\phi$ (as shown by \ref{eq:interface_equation}).
In the following, we thus take $\phi= 1$ in \ref{eq:dl_mc_RaPr}; the only effect of this is to change the regularisation length at the interface, a regularisation that must anyway be smaller than any dynamical lengths for the model to remain valid, as discussed in more detail in \ref{sec:numerics}.
%
%

%
In the heat transport equation \ref{eq:dl_ec_RaPr}, we reduce the number of control parameters by taking $\overline{\phi} = \Lambda = 1$.
For hydrothermal systems, water flows through a silicate rock matrix.
The thermal diffusivity is typically a factor of two larger in the matrix than in the fluid, while $\rho_m c_m \sim \rho c$. The parameters 
$\overline{\phi}$ and $\Lambda$ are thus order one constants that do not vary substantially from one system to another. 
This is perhaps less true in some industrial applications like the transport of heat through the metallic foam \citep{su_macroscopic_2015} where the thermal conductivity can be at least a hundred times larger in the matrix than in the fluid. This would lead to a large value of $\Lambda$ and thermal diffusion would be greatly enhanced in the porous medium, reducing its ability to convect. 
We do not consider such cases here, although we will find that cases where the porous medium is dominated by diffusive transport can be easily treated theoretically, and the theory could be straightforwardly adapted to account for varying $\Lambda$. 
Finally, to reduce the complexity of this study and maintain a focus on the key features of varying the driving buoyancy forces (i.e. $\Ra$) and the properties of the porous matrix (i.e. $\Da$), we set the Prandtl number to be $\Pr = 1$ throughout this work. 

\subsection{Limits on the control parameters}
\label{sec:limits}

Several constraints must be imposed on the control parameters $\Ra$ and $\Da$ to ensure that the Darcy-Brinkman model remains valid. We give these constraints in their most general form here, but recall from the previous section that all solutions in this work take $\Pr = \Lambda = 1$. 
First, the inertial terms must vanish in the porous medium, which demands that 
\begin{equation}
\frac{\sqrt{\Ra} \Da}{\sqrt{\Pr}} \ll 1;
\label{constraint1}
\end{equation}
that is, the velocity scale in the porous medium must be much less than the $O(1)$ velocity in the unconfined fluid layer. 

Second, the continuum approximation that underlies Darcy's law requires that any dynamical lengthscale of the flow in the porous layer must be larger than the pore scale $\sqrt{\Da}$; equivalently, the Darcy drag term must always be larger than local viscous forces in the bulk of the medium.  
We expect the smallest dynamical scales to arise from a balance between advection and diffusion in \ref{eq:dl_ec_RaPr}: such a balance, given typical velocity $\sim \RaDa/\sqrt{\Pr}$, yields a lengthscale $\sim \Ra_p^{-1}$. In fact, simulations carried out in a porous Rayleigh-B\'enard cell \citep{hewitt_ultimate_2012} indicate that the narrowest structures of the flow, which are thermal boundary layers, have a thickness of at least $50 \Ra_p^{-1}$. For these structures to remain larger than the pore scale $\sqrt{\Da}$, we thus require 
\begin{equation}
\label{eq:limit_pore_scale}
\Ra \Da^{3/2} \lesssim 50 \Lambda.
\end{equation}
Note that the effect of violating this constraint is to amplify the importance of viscous resistance $\nabla^2 \boldsymbol{u}$ within the porous medium in \ref{eq:dl_mc_RaPr}, which would no longer reduce to Darcy's law.

%

\begin{figure}
\centerline{
\includegraphics[width=0.6\linewidth]{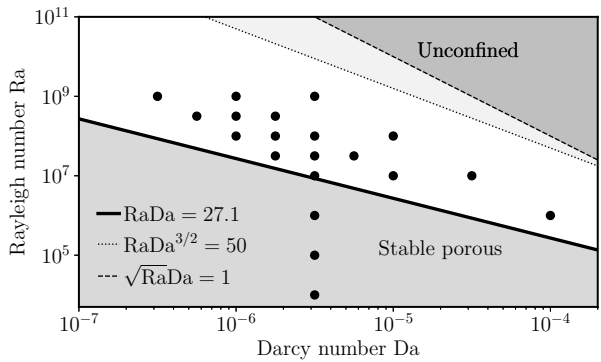}
}
\caption{Domain of existence and validity of the two-layer model with respect to the control parameters $\Ra$ and $\Da$, given $\Pr = \Lambda = 1$. Each dot represent a simulation. The line $\RaDa= 1$ marks the limit beyond which inertial terms affect the flow in the porous medium, while the line $\Ra \Da^{3/2} = 50$ gives an estimate of the point at which the smallest flow structures in the porous medium become comparable to the pore scale. The line $\Ra \Da = \Ra_p^c =  27.1$ corresponds to the threshold of thermal convection in a porous \RB cell with an open-top boundary as discussed in \ref{sec:porous_stability}.}
\label{fig:all_simulations}
\end{figure}

%
\Cref{fig:all_simulations} provides an overview of the space of control parameters $\Ra$ and $\Da$ where these various limits are identified and the parameter values in our numerical simulations are indicated. 
This plot also shows a line that approximately marks the threshold of convective instability in the porous medium, whose importance is discussed in \ref{sec:overview_heat_transport} and which is theoretically quantified in \ref{sec:porous_diffusive_regime}.

\subsection{Numerical method}
\label{sec:numerics}

The one-domain Darcy-Brinkman equations \ref{eq:dimensionless_equations} are solved using the pseudo-spectral code \textsc{Dedalus} \citep{burns_dedalus_2020,hester_improving_2019}.
The flow is decomposed into $N$ Fourier modes in the horizontal direction, while a Chebyshev polynomial decomposition is used in the vertical direction.
Because the set-up is comprised of two layers whose interface must be accurately resolved, each layer is discretised with its own Chebyshev grid, with $[M_p,M_f]$ nodes the porous and fluid layers, respectively.  
This ensures enhanced resolution close to the top and bottom boundary as well as at the interface where the $\sqrt{\Da}$ regularisation length must be resolved. 
Time evolution of the fields is computed with implicit-explicit methods \citep{wang_variable_2008}: the non-linear and Darcy terms in \ref{eq:dimensionless_equations} are treated explicitly while viscosity and diffusion are treated implicitly.
The numerical scheme for time evolution uses second order backward differentiation for the implicit part and second order extrapolation for the explicit part \citep{wang_variable_2008}.
The stability of temporal differentiation is ensured via a standard CFL criterion \textcolor{RED}{evaluating the limiting timestep in the whoe two-layer domain}, with an upper limit given by  $\sqrt{\Ra} \Da$, which is never reached in practice.
Rather, \textcolor{RED}{with the control parameters and resolution considered here}, the time step is limited by the non-zero vertical velocity at the $z=0$ interface where the vertical discretisation is refined. 
The range of Rayleigh and Darcy numbers in our simulations is shown in \ref{fig:all_simulations}. Note, with reference to this figure, that we carried out a systematic investigation of parameter space where the porous layer is unstable by varying $\Ra$ and $\Da$ for various fixed values of the porous velocity scale $\RaDa$. 
%

%
%
%

%
The majority of simulations were carried out in a set-up where the heights of both the porous and the fluid layer were equal $\hat{h}_p = \hat{h}_f = 1$, with resolutions $N \times [M_p,M_f] = 1024 \times [128, 256]$ below $\Ra = 10^{8}$ and $1024 \times [256, 512]$ above.
As discussed in the following sections, we find that the porous layer in general absorbs more than half of the temperature difference, and so the effective Rayleigh number in the fluid layer is typically somewhat smaller than $\Ra$. 
%
%
%

We used two methods to initiate the simulations. 
In a few cases, the initial condition was simply taken as the diffusive equilibrium state throughout the domain, perturbed by a small noise.  
In most cases, however, we proceeded by continuation, using the final output of a previous simulation similar control parameters as an initial condition for a new simulation. 
%
%
Comparison between the two methods showed that they yielded the same statistically steady state, but the continuation approach reached it in the shortest time. In all cases, we ran simulations over a time comparable to the diffusive timescale $\sqrt{\Ra}$, to ensure that the flow had reached a statistically steady state. 
%

%
%

%

\section{An overview of two-layer convection}
\label{sec:overview}

In this section we describe the results of a series of simulations carried out at a fixed Darcy number, $\Da = 10^{-5.5}$, and equal layer depths $\hat{h}_p = \hat{h}_f = 1$, but with varying $\Ra$ in the range $10^4 \leqslant \Ra \leqslant 10^9$.
We use these to illustrate the basic features of high-$\Ra$ convective flow in the two-layer system. 

\subsection{Two different regimes depending on the stability of the porous medium}
\label{sec:porous_stability}

\begin{figure}
\includegraphics[width=\linewidth]{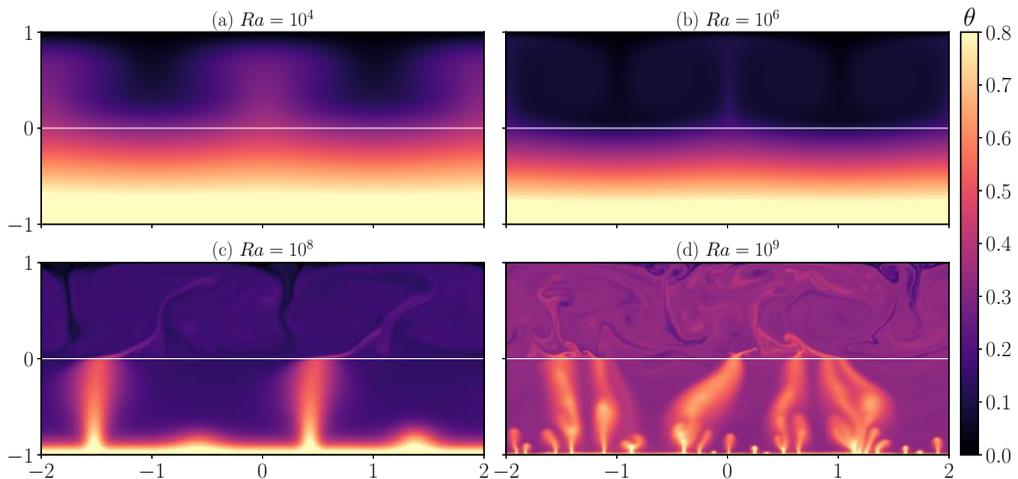}
\caption{Snapshots of the temperature field at different Rayleigh numbers in two stable (top panels) and two convective (bottom panels) cases for the porous medium. The Darcy number is kept at $\Da = 10^{-5.5}$.
The colour scale is cut at $0.8$ to enhance the contrast in the fluid layer. 
 }
\label{fig:snapshots_Da1e_5_5}
\end{figure}

\Cref{fig:snapshots_Da1e_5_5} shows snapshots of the temperature field taken for different simulations that have reached a statistically steady state. 
The corresponding profiles of the horizontally and temporally averaged temperature, $\overline{\theta}(z)$, are shown in \ref{fig:profiles_Da55}(a), while the mean interface temperature $\theta_i = \overline{\theta}(0)$ is shown in \ref{fig:profiles_Da55}(b).
The fluid layer is convecting in all simulations, as attested by the presence of plumes and by the mixing of the temperature field that tends to create well-mixed profiles of $\overline{\theta}$ in the bulk of the fluid.
The porous layer, however, exhibits two different behaviours depending on the size of $\Ra$. 
At low Rayleigh numbers ($\Ra \leq10^6$ in \ref{fig:snapshots_Da1e_5_5}), the porous layer is dominated by diffusive heat transport: there are no hot or cold plumes in the temperature field in $z < 0$ (see \ref{fig:snapshots_Da1e_5_5}a,b), while the horizontally averaged temperature profiles $\overline{\theta}(z)$ appear to be linear (\ref{fig:profiles_Da55}a).  The corresponding interface temperature monotonically decreases with $\Ra$ (\ref{fig:profiles_Da55}b)  
As the Rayleigh number is increased beyond $\Ra \sim 10^7$, the behaviour of the flow in the porous layer changes as it also becomes unstable to convection.
This is attested by the visible presence of plumes in \ref{fig:snapshots_Da1e_5_5}(c,d) and by the flattening of the horizontally averaged temperature profiles in \ref{fig:profiles_Da55}(a). 
The signature of this transition is also very clear in the evolution of the interface temperature $\theta_i$, which reverses from decreasing to increasing with $\Ra$ around $\Ra \sim 10^7$ (\ref{fig:profiles_Da55}b).

The value of the Rayleigh number at which porous convection emerges can be roughly estimated from the stability of a single porous layer.
\textcolor{RED}{
In a standard Rayleigh-B\'enard cell with open-top boundary, convection occurs if the porous Rayleigh number $\Ra_p = \Ra\Da$ exceeds a critical value $\Ra_p^c \simeq 27.1 $ \citep{nield_convection_2017}.
}
At $\Da = 10^{-5.5}$, the critical Rayleigh number $\Ra$ such that $\Ra\Da = \Ra_p^c$ is $\Ra \simeq 8.6 \times 10^{6}$, which is reported in \ref{fig:profiles_Da55}(b) and agrees well with the inversion of trend in $\theta_i$.  We will return to a more nuanced form of this argument in \ref{sec:porous_diffusive_regime}. 
\textcolor{RED}{
Lastly, as the Rayleigh number is increased beyond the threshold of porous convection, the porous plumes become thinner and more numerous, a behaviour that is similar to standard Rayleigh-B\'enard convection in porous media \citep{hewitt_ultimate_2012}.
In addition, the porous plumes become increasingly narrower at their roots in the thermal boundary layer, which causes a local minimum in the temperature profiles (see figure \ref{fig:profiles_Da55}a) that has also been observed in previous works on porous convection at large Rayleigh number \cite{hewitt_ultimate_2012}. 
Lastly, the temperature contrast across the interface reduces as the Rayleigh number is increased. 
}

\subsection{Characteristics of heat transport}
\label{sec:overview_heat_transport}

\begin{figure}
\centerline{
\includegraphics[width=\linewidth]{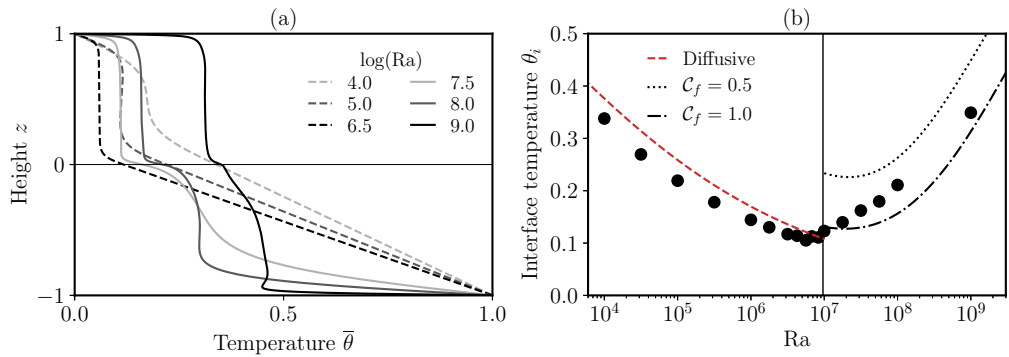}
}
\caption{Temporally averaged quantities for $\Da = 10^{-5.5}$. \textbf{(a)} Horizontally averaged temperature $\overline{\theta}(z)$ for values of $\Ra$ below (dashed) and above (solid) the threshold of porous convection; and  
\textbf{(b)} mean interface temperature $\theta_i$.  The red dashed line is the diffusive prediction of \ref{eq:diffusive_interface_temperature} and the black lines are asymptotic predictions obtained by solving \ref{eq:interface_temperature_equation} using $\mathcal{C}_{p} = 0.85$ and the marked values $\Cf$, as detailed in section \ref{sec:interface_temperature}. 
}
\label{fig:profiles_Da55}
\end{figure}

\begin{figure}
\includegraphics[width=\linewidth]{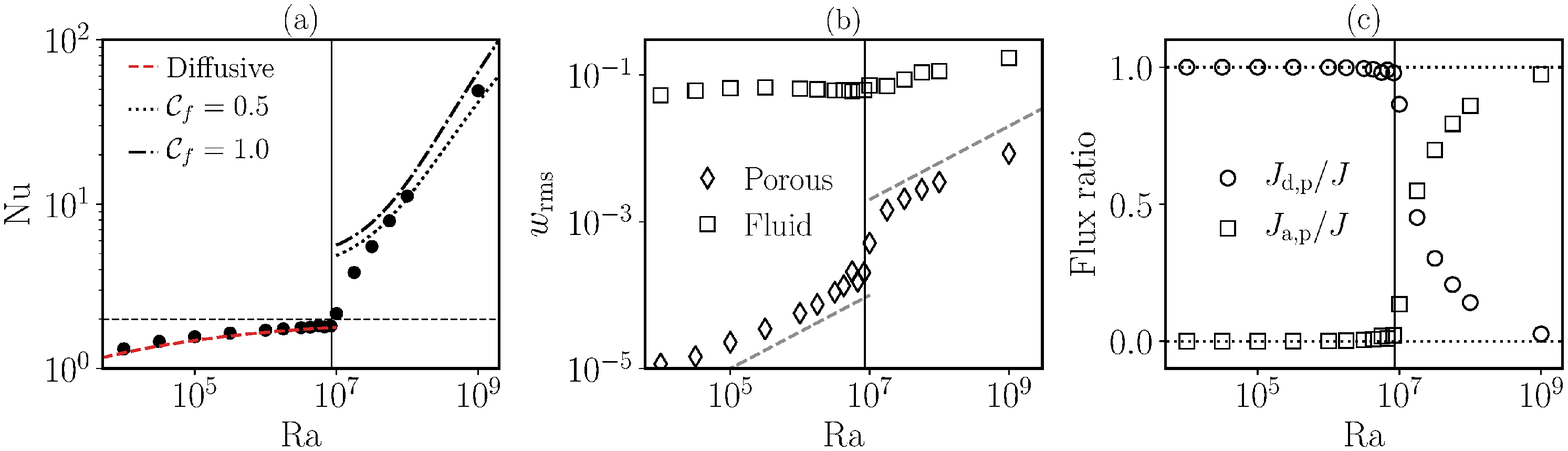}
\caption{
\textbf{a:} Nusselt number $\Nu = 2 \sqrt{\Ra} \left<J\right>$ characterising heat transport across the two-layer system, together with predictions from \ref{sec_theory} for the porous-stable case \ref{eq:diffusive_interface_temperature} (red dashed) and porous-unstable \ref{eq:interface_temperature_equation} (black dotted and dot-dashed, with  $\mathcal{C}_{p} = 0.85$ and $\mathcal{C}_{f}$ as marked). 
The horizontal line marks $\Nu = 2$.
%
%
\textbf{b:} Root-mean-squared vertical velocity amplitude $\wrms$ in the porous and fluid layers. The two grey lines indicate a scaling of $\RaDa$ (the characteristic speed in the porous layer).  
\textbf{c:} Ratio of the diffusive ($J_{\mathrm{d,p}}$) and advective ($J_{\mathrm{a,p}}$) fluxes to the total depth-averaged flux $\Jav$ in the porous layer. 
In all three panels, $\Da = 10^{-5.5}$ and the vertical line marks the threshold of porous convection estimated using \ref{eq:exact_porous_threshold}. 
}
\label{fig:global_Da55}
\end{figure}
%

%
The transition between the porous stable and the porous unstable case can be further identified by the analysis of global heat transport across the system. 
Heat transport is characterised by the horizontally averaged heat flux $J(z) \equiv \overline{w \theta} - \Ra^{-1/2} \overline{\theta}' $, which is constant with height in a statistically steady state.
As is standard in statistically steady convection problems, we measure the time-averaged enhancement of the heat flux, compared to what it would be in a purely diffusive system, $\Ra^{-1/2} /2$, via the Nusselt number $\Nu \equiv 2 \sqrt{\Ra} \left<J\right>$, where the angle brackets indicate a long-time average. 
The Nusselt number (\ref{fig:global_Da55}a) is strongly influenced by the transition to instability in the porous layer. 
In the porous-stable case ($\Ra \lesssim 10^7$), $\Nu$ appears to approach a horizontal asymptote $\Nu = 2$, but once the porous layer is unstable $\Nu$ increases much more steeply beyond this value.

The behaviour below the threshold of convection is due to the flux being predominantly diffusive in the porous layer. 
The total flux through the system is thus bounded above by a state in which all of the temperature contrast is taken up across the porous layer and the interface temperature $\theta_i$ tends to zero. In this limit, $\Ja \to 1/\sqrt{\Ra}$ and $\Nu \to 2$.  The decreasing $\theta_i$ in \ref{fig:profiles_Da55}(b) as $Ra$ increases towards the threshold reflects the approach to this limit.
In fact, we find that despite the porous medium appearing to be stable to convection below the threshold, small amplitude flows still exist in this regime, as can be seen from the non-zero root-mean-squared vertical velocity in the porous layer for all $Ra$, shown in \ref{fig:global_Da55}(b). 
Nevertheless, by computing the relative diffusive and advective contributions to the flux through the porous medium, we confirm that these flows have a negligible impact on heat transport below the threshold (\ref{fig:global_Da55}c).
We interpret the weak secondary porous flows in the porous-stable regime as a consequence of the horizontal variations in the interface temperature imposed by fluid convection, which are clearly visible in \ref{fig:snapshots_Da1e_5_5}(a,b). 
As shown in \ref{fig:global_Da55}(b,c), the strength of the porous flow increases dramatically as the porous layer becomes unstable, and it is only then that the advective contribution to the flux in the porous medium becomes significant.  We return to discuss this induced flow below onset in the porous layer in \ref{penetrative_sec}, and defer more detailed discussion and prediction of the behaviour of $\Nu(\Ra,\Da)$ until the following section.

\subsection{Timescales, variability and statistically steady state}
\begin{figure}
\centerline{
\includegraphics[width=0.90\linewidth]{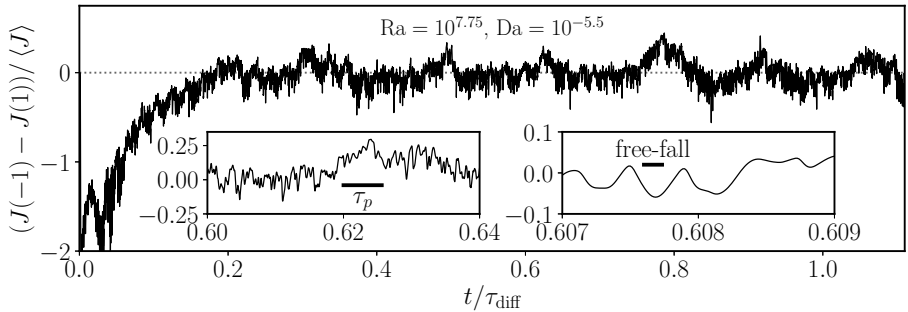}
}
\caption{
\textbf{a:} Time series of the flux difference between top and bottom at $\Ra = 10^{7.75}$, $\Da = 10^{-5.5}$.
The flux difference is normalised by the volume and temporal average of the flux $\Ja$ and time is normalised by $\taud = \sqrt{\Ra}$.
\textcolor{RED}{
The insets show zooms to make clearer the different levels of temporal variations, with horizontal segments indicating the porous turnover timescale $\tau_p$ and the free-fall timescale. 
}
Note that this simulation has been started from a diffusive state with a small noise.
}
\label{fig:heat_conservation}
\end{figure}

The governing equations \ref{eq:dimensionless_equations} reveal three different timescales that govern the variability of the two-layer system considered here. 
The first is the turnover time scale in the fluid layer, which is $O(1)$ in our free-fall normalisation.
%
%
The second is given by diffusion, $\taud = \sqrt{\Ra}$, and
the third is the turnover timescale in the porous layer $\tau_p \sim (\RaDa)^{-1}$, which scales with the inverse of the porous speed scale. 
Because $\tau_p$ and $\tau_d$ measure advection and diffusion in the porous medium, these timescales are in a ratio $\tau_p = \taud/\Ra_p$.
In addition, we recall that $\tau_p \gg 1$ is required for the porous layer to be in the confined limit and for the Darcy--Brinkman model to hold (see \ref{constraint1}).
The turnover timescales should scale with the inverse of $\wrms$ in each layer, as can be observed in \ref{fig:global_Da55}(b): in the fluid layer, $\wrms \sim O(1)$ with no systematic variation with $Ra$, while in the porous layer, $\wrms \propto \sqrt{\Ra}$ at constant $\Da$ in agreement with the scaling above. 
These two very different timescales are also clearly visible in time series of the heat flux difference across the two-layer system, as shown in \ref{fig:heat_conservation}.  Fast oscillations driven by the fluid convection variability are superimposed onto longer time variations due to flow in the porous layer. 
Such a time series also illustrate how the two-layer set-up reaches a statistically steady state, the latter being characterised by the flux difference averaging to $0$ over long times.
In the particular case of \ref{fig:heat_conservation}, the simulation is initiated from the diffusive equilibrium plus a small noise and we observe the equilibration to occur after $\sim 0.2 \taud$. 
Although the equilibration time is largely reduced by the use of continuation, we run all simulations over times that are similar to $\taud$ to ensure they are converged.

\section{Modelling heat transport and interface temperature}
\label{sec_theory}

%

\subsection{Predicting heat transport from individual layer behaviour}

As observed in \ref{fig:profiles_Da55}(a), when both layers are convecting the temperature profiles in each layer are characterised by thin boundary layers at either the upper or lower boundaries of the domain, through which heat must diffuse. This structure is a generic feature of convection problems, and suggests that we may be able to generalise previous results and approaches used in standard Rayleigh--B\'enard convection problems in order to predict the behaviour here. 

%

\subsubsection{An asymptotic approach based on boundary-layer marginal stability}
\label{sec:asymptotic_BL}

\begin{figure}
\centerline{
\includegraphics[width=0.6\linewidth]{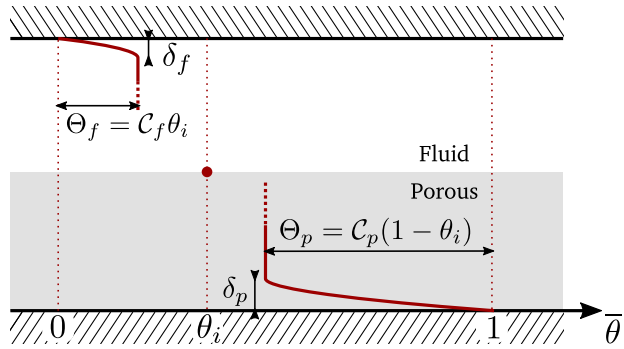}
}
\caption{A schematic of the temporal and horizontal average of the temperature in the two-layer system, with a focus on the fluid and porous boundary layers. }
\label{fig:boundary_layers}
\end{figure}

Following the seminal approach of \cite{malkus_heat_1954} and \cite{howard_convection_1966}, we posit that in the asymptotic limit of large Rayleigh and small Darcy numbers such that $\RaDa \ll 1$, the thermal boundary layers at the upper and lower boundaries of the domain are held at a thickness that is marginally stable to convection. 
In order to apply this idea, let us consider a general region with a Rayleigh number $R$ (i.e. $\R = \Ra$ in the fluid layer and $\R = \Ra \Da$ in the porous layer). If the boundary layer has mean thickness $\delta$, then we can also introduce a local boundary-layer Rayleigh number $R \delta^\beta$, where $\beta =1$ in the porous layer or $\beta =3$ in the fluid layer to account for the different dependence on the height scale in the corresponding Rayleigh number (see \ref{parameters} and \ref{Rap}). Let the temperature difference across the boundary layer be $\Theta$ (\ref{fig:boundary_layers}), and, for completeness, suppose that the region has an arbitrary dimensionless height $\hh$. 

Suppose further that we want to rescale lengths and temperatures to compare this general case more directly with the standard Rayleigh--B\'enard cell of unit dimensionless height and unit temperature difference. In such a cell, the temperature contrast across the boundary layers is $1/2$, which suggests that we need to rescale temperatures by $2 \Theta$ and lengths by $\hh$, to give a new Rayleigh number $\hR = 2\Theta \hh \R$ and boundary-layer depth $\hat{\delta} = \delta/\hh$. Having rescaled thus, the Malkus--Howard approach would suggest that
\begin{equation}
\label{eq:marginal_stabilty_BL}
\hR \hat{\delta}^\beta = \R_c,
\end{equation}
for some critical value $\R_c$, or $\hat\delta = (\R_c/\hR)^{1/\beta}$. The corresponding Nusselt number $\N$ for this rescaled system, given that the scaled temperature drop across the layer is $1/2$, is
\begin{equation}
\label{eq:rescaled_Rayleigh_Nusselt}
\N = \frac{1}{2\delta} = \frac{1}{2} \left(\frac{\hR}{\R_c}\right)^{1/\beta}.
\end{equation}
Note that the actual, unscaled flux $\left<J\right>$ across the boundary layer is $\left< J \right> = \Ra^{-1/2} \Theta/\delta$, which can thus be related to $\N$ from \ref{eq:marginal_stabilty_BL} and \ref{eq:rescaled_Rayleigh_Nusselt} by
\begin{equation}
\label{eq:orig_flux}
\left<J \right> = \frac{ 2 \Theta}{\hh \sqrt{\Ra}}  \N.
\end{equation} 

Thus specification of the critical value $\R_c$ yields a prediction of the flux through the system in terms of $\Theta$ and $\hh$. We can extract values of $\R_c$ from previous works that have determined experimentally or numerically the relation between $\N$ and $\hR$ in either a fluid or a porous \RB~cell.
For porous convection, \cite{hewitt_ultimate_2012} found that $(2 \R_c^{1/\beta})^{-1} \simeq 6.9 \times 10^{-3}$. 
For unconfined fluid convection, the host of historical experiments reported in \cite{ahlers_heat_2009} and \cite{plumley_scaling_2019}, as well as more recent studies for instance by \cite{urban_heat_2014} or \cite{cheng_laboratory-numerical_2015}, suggest that $(2 \R_c^{1/\beta})^{-1} \sim 6-8 \times 10^{-2}$, although no definitive observation of a well-developed $\Ra^{1/3}$ law has been made so far and the `true' asymptotic form of $\N(\hR)$ remains a hotly contested questions.
Nevertheless, these values provide an estimate for the heat flux in both the porous and the fluid layer in the asymptotic limit $\Ra \gg 1$ and $\RaDa \ll 1$ that can be compared to our simulations, as discussed in the next section. 
\subsubsection{Generalising the flux estimate using Rayleigh-Nusselt laws}

Our simulations remain limited to moderate Rayleigh numbers, mainly because of the flows through the interface that need to be accurately resolved, and so the asymptotic arguments outlined in the previous section may not be accurate. 
%
%
However, it is straightforward to generalise the asymptotic approach of the previous section to a case where the Nusselt number is some general function $\mathcal{N}(\hR)$ of the rescaled Rayleigh number, rather than the asymptotic scaling. To achieve this, we simply replace \ref{eq:rescaled_Rayleigh_Nusselt} by the relationship 
\begin{equation}
\label{eq:general_nu}
\N = \mathcal{N}(\hat{R}).
\end{equation}
(Equivalently, one could generalise the asymptotic results above to allow $R_c$ to vary with $\hat{R}$ in the manner necessary to recover \ref{eq:general_nu}.)  Over the range of fluid Rayleigh numbers considered here, \cite{cheng_laboratory-numerical_2015} determined an approximate fit to the fluid Nusselt number function $\mathcal{N}_f$ of
\begin{equation}
\label{eq:fluid_Nusselt}
\mathcal{N}_f (\hR_f) = 0.162\, \hR_f^{0.284}.
\end{equation} 
In the porous layer, \cite{hewitt_ultimate_2012} and \cite{hewitt_vigorous_2020} considered the equivalent porous Nusselt number $\mathcal{N}_p$ and found a correction to the asymptotic relationship of the form $\mathcal{N}_p (\hR_p) = 6.9 \times 10^ {-3}\, \hR_p + 2.75$, which recovers the asymptotic relationship in the limit $\hR_p \rightarrow \infty$. In fact, we also carried out additional simulations of porous convection in a layer of height $\hh = 1$ submitted to a maintained temperature difference of $1$, but in which the top boundary is open as it allows the fluid to flow in and out. 
An affine fit of the Nusselt number against the porous Rayleigh number in these simulations provided us with the law
\begin{equation}
\label{eq:porous_Nusselt}
\mathcal{N}_p (\hR_p) = 6.99 \times 10^ {-3}\, \hR_p + 1.56~.
\end{equation}
Note that in such a cell, the temperature difference across the bottom boundary layer is not $1/2$ as in a classical Rayleigh-B\'enard cell, but rather about $\Theta \sim 0.8$.
Note also that the linear coefficient of the fit in \ref{eq:porous_Nusselt} is effectively the same as that found in the classical cell, which indicates that for sufficiently large porous Rayleigh number the mechanisms controlling the lower boundary layer are the same regardless of the nature of the interface condition. 
\begin{figure}
\includegraphics[width=\linewidth]{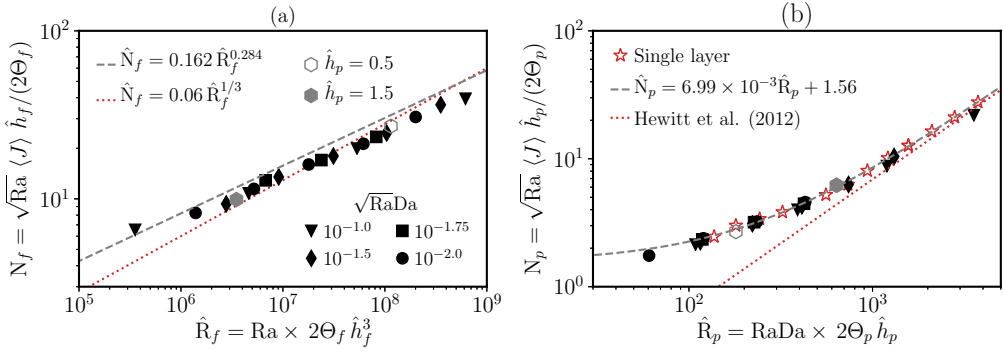}
\caption{
The Nusselt number $N$, extracted from the measured flux $\left<J\right>$ and temperature contrast across the boundary layers via \ref{eq:orig_flux}, as a function of the rescaled Rayleigh number $\hR = \R\,2 \Theta \hh^\beta$ in the fluid layer (panel \textbf{a} and subscripts $f$) and in the porous  layer (panel \textbf{b} and subscripts $p$).
\textcolor{RED}{
The values of $\Theta$, the temperature difference across the boundary layers, is extracted from numerical data. 
}
Numerical data are sorted by equal values of the damping factor $\RaDa$. 
In each panel, fits from studies of the single-layer case \ref{eq:fluid_Nusselt} or \ref{eq:porous_Nusselt} are included (dashed grey) together with asymptotic predictions for $\hR \to \infty$ as discussed in \ref{sec:asymptotic_BL} (dotted red). Panel \textbf{b} also includes data from simulations of a single porous Rayleigh-B\'enard cell with an open top boundary on which the fit \ref{eq:porous_Nusselt} is based (red stars).  All but two simulations had $\hh_p = \hh_f = 1$; the values of $\hh_p$ for these two simulations are given in the legend. 
}
\label{fig:Ra_Nu_fluid}
\end{figure}
\Cref{fig:Ra_Nu_fluid} shows a comparison of the predictions of this theory and various flux laws with our numerical data. 
We show the Nusselt number $\N$, calculated in our simulations using \ref{eq:orig_flux}, as a function of the rescaled Rayleigh number $\hR$ in both the fluid layer (index $f$) and in the porous layer (index $p$), \textcolor{RED}{with the temperature drop across boundary layers $\Theta$ determined from numerical simulations.}
We have indicated for reference the single-layer laws \ref{eq:fluid_Nusselt} and \ref{eq:porous_Nusselt} which are in close agreement with our data. We also include two simulations with different layers depths $\hh_p$ and $\hh_f$ to demonstrate the generality of the theory. 
In \ref{fig:Ra_Nu_fluid}(a), we observe that the points lie slightly below the law \ref{eq:fluid_Nusselt} found by \cite{cheng_laboratory-numerical_2015}, but we note that our values of $\N_f$ lie within the range of the scatter in the data upon which that fit is based in \cite{cheng_laboratory-numerical_2015}. 
%
%
%
Overall, the figure indicates that flux laws extracted from individual layers can be used to predict the flux in the two-layer system, after careful rescaling.

%
%

%
%
\subsection{The interface temperature}
\label{sec:interface_temperature}

\begin{figure}
\includegraphics[width=\linewidth]{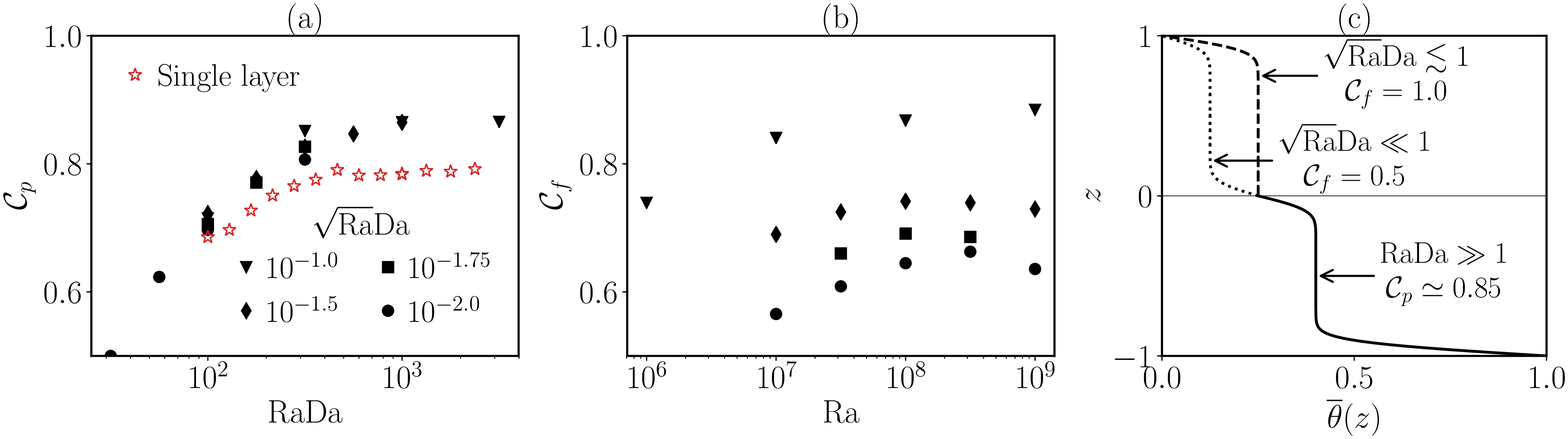}
\caption{
\textbf{a \& b:} Values of $\Cp$ and $\Cf$ (see \ref{fig:boundary_layers}) extracted from simulations for different porous velocity scale $\RaDa$. 
\textbf{c:} Schematic representation of two limiting temperature profiles in the limit of large Rayleigh numbers based on panel a and b. The porous Rayleigh number is large so that $\Cp = 0.85$ whereas $\Cf$ varies from $1$ in the weakly confined limit ($\RaDa \lesssim 1$) to $0.5$ in the strongly confined limit ($\RaDa \ll 1$).
}
\label{fig:CpCf_values}
\end{figure}

We can use the laws governing the heat flux to determine the interface temperature $\theta_i$, 
which is important for applications as it gives an order of magnitude of the average fluid temperature. 
The interface temperature must be set by the constraint of flux conservation between the two layers, which implies that the flux in \ref{eq:orig_flux} must be the same in both porous and fluid layers, in a statistically steady state. 
The caveat is that it is not clear how to relate the temperature difference $\Theta$ across the boundary layers to the interface temperature $\theta_i$. 
We overcome this issue by saying that the temperature difference across the boundary layer is a fraction of the difference across the whole layer, and so introduce $O(1)$ coefficients $\mathcal{C}_{f,p}$ such that $\Theta_{f} = \Cf \theta_i$ and $\Theta_p = \Cp (1-\theta_i)$ (see \ref{fig:boundary_layers}).
In single classical Rayleigh-B\'enard cells, $\mathcal{C}_{f,p} = 0.5$ because both boundary layers are symmetric and diffusive.
This is not true in the two-layer system where the transport across the interface includes contributions from diffusion and advecion, and $\mathcal{C}_{f,p}$ may take any value ranging from $0.5$ to $1$.

Given these coefficients, flux conservation between the layers yields
\begin{equation}
\label{eq:interface_temperature_equation}
\frac{2 \mathcal{C}_f \theta_i}{ \hh_f} \mathcal{N}_f (2 \mathcal{C}_f \theta_i \hh_f^3 \Ra) =  \frac{2 \mathcal{C}_p (1-\theta_i)}{\hh_p} \mathcal{N}_p (2 \mathcal{C}_p (1-\theta_i) \hh_p \Ra\Da),
\end{equation}
which determines the interface temperature $\theta_i$.  
In general, the fractions $\mathcal{C}_{f,p}$ depend on the control parameters $\Ra$ and $\Da$, as shown in \ref{fig:CpCf_values}.

\subsubsection{The porous-convective regime}
\begin{figure}
\centerline{
\includegraphics[width=\linewidth]{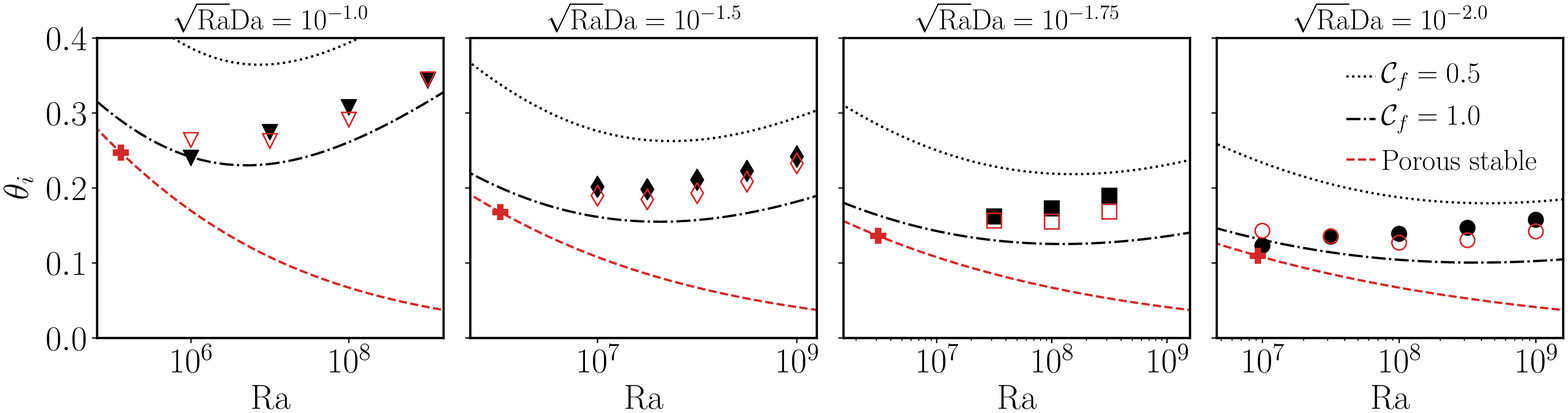}
}
\caption{
Plot of the interface temperature $\theta_i$ as a function of the Rayleigh number $\Ra$ for different values of $\RaDa$. 
The filled symbols are the numerical data and the empty symbol represent is the predicted $\theta_i$ using equation \ref{eq:interface_temperature_equation} with the values of $\mathcal{C}_{p,f}$ extracted from the simulation and shown in \ref{fig:CpCf_values}a \& b.
The black lines show the law expected for the interface temperature in the two limiting cases of strongly confined porous medium ($\RaDa \rightarrow 0$, $\Cp = 0.85$ and $\Cf = 0.5$) and the weakly confined case ($\RaDa \lesssim 1$, $\Cp = 0.85$ and $\Cf = 1.0$).
The red line is the interface temperature in the case where the porous medium remains diffusive, the plus sign highlighting the value at threshold of porous convection based on the prediction \ref{eq:exact_porous_threshold}. 
}
\label{fig:interface_temperature}
\end{figure}
We first address the case where both layers are unstable to convection.
\textcolor{RED}{
Although $\Cp$ and $\Cf$ may take any value between $0.5$ and $1$, we can still make qualitative predictions on their values depending on the control parameters.
Because these coefficients describe the temperature difference between the bulk and the boundary layers, they are also a proxy for the temperature drop at the interface.
In the case where the porous turnover timescale $\tau_p = \RaDa$ is very small compared to the free-fall timescale ($= 1$ in dimensionless units), the porous flow is strongly confined and heat transfer is purely diffusive at the interface so that $\Cp = \Cf = 0.5$.
In the opposite, weakly confined limit where $\RaDa$ is brought up to 1, the porous and fluid velocities become similar and the interface heat transfer becomes advective which forces the interface temperature drop to vanish (see for instance the $\Ra = 10^9$ temperature profile in \ref{fig:profiles_Da55}a for which $\RaDa=10^{-1}$).
In such a limit, $\Cp = \Cf = 1$.
}
\textcolor{RED}{
The values of $\Cp$ and $\Cf$ extracted from numerical simulations are shown in \ref{fig:CpCf_values}a and b, respectively. 
The qualitative picture described above seems in agreement with the numerical observation in the fluid coefficient: as shown in \ref{fig:CpCf_values}b, although $\Cf$ varies with $\Ra$ it is also larger for larger values of $\RaDa$.
Given there is not a clear collapse of this data, we consider in the following discussion on the interface temperature two extreme limits in the asymptotic limit $\Ra \to \infty$, as sketched in \ref{fig:CpCf_values}(c), $\Cf = 0.5$ in the strongly confined limit ($\RaDa \ll 1$ ) and $\Cf = 1$ in the weakly confined limit ($\RaDa \lesssim 1$). 
}

\textcolor{RED}{
In the porous layer, the numerical data shows difference with the qualitative expactations for the values of $\Cp$ (see \ref{fig:CpCf_values}a).
$\Cp$ is mainly a function of the porous Rayleigh number $\Ra_p = \Ra \Da$ (\ref{fig:CpCf_values}a), and appears to approach an asymptote $\Cp \simeq 0.85$. 
Interestingly, the trend of $\Cp$ with $\Ra_p$ is the same as in a single porous layer with an open top (red stars in \ref{fig:CpCf_values}a), but with smaller values and lower asymptote in that case ($\simeq 0.78$). 
In the confined porous limit, this is due to the temperature drop at the interface being shared by both the fluid and the porous layer.
In the weakly confined limit where the interface temperature drip tends to vanish, the value of $\Cp \simeq 0.85$ rather reflects the temperature decrease in the bulk of the porous layer (see the $\Ra = 10^9$ temperature profile in \ref{fig:profiles_Da55}a).
}
%

%
Before showing asymptotic predictions based on these limiting values of $\mathcal{C}_{p,f}$, we first use the actual extracted values to verify the accuracy of \ref{eq:interface_temperature_equation} in predicting the interface temperature (\ref{fig:interface_temperature}). We find the agreement between both to be within $10$ \% relative error.
This figure also shows the predicted interface temperature for $\Ra \gg 1$ for the two limiting cases $\Cf = 0.5$ and $\Cf = 1.0$ with $\Cp = 0.85$ as discussed above.
Apart from values close to the threshold of porous convection, all the numerical data lies between them.
Moreover, the weakly confined simulations with larger $\RaDa$ show interface temperatures approaching the limit $\Cf =1$ (\ref{fig:interface_temperature}a), while the more strongly confined simulations with $\RaDa \ll 1$ draw closer to $\Cf = 0.5$ as $\Ra$ is increased (\ref{fig:interface_temperature}d), which suggests that these limits will become increasingly accurate for increasingly large $Ra$. 
%
%

%
\subsubsection{The porous-diffusive regime}
\label{sec:porous_diffusive_regime}

If the porous layer remains stable, dominated by diffusive heat transfer, then the expression for the flux in the fluid layer is still given by \ref{eq:orig_flux}, but in the porous layer it is simply $\left< J \right> = \Ra^{-1/2} (1-\theta_i)/\hh_p$. A balance of these expressions should then replace \ref{eq:interface_temperature_equation} to determine $\theta_i$. Moreover, we know \textit{a priori} that $\mathcal{C}_{f} = 0.5$ since the flux is entirely diffusive at the interface $z=0$. Flux conservation thus reduces to 
\begin{equation}
\label{eq:diffusive_interface_temperature}
\frac{\theta_i}{\hf} \mathcal{N}_f(\theta_i \hh_f^3\Ra ) = \frac{1-\theta_i}{\hp},
\end{equation} 
which we solve numerically. The predicted interface temperature from this model was shown in \ref{fig:profiles_Da55}(b) and gives reasonably good agreement with the numerical data. 

Given $\theta_i$, we can also extract the Nusselt number for the two-layer system, $\Nu = 2 \sqrt{\Ra} \left<J\right> = 2 (1-\theta_i)/ \hat{h}_p$ (see \ref{sec:overview}), which was also shown giving good agreement with numerical data in \ref{fig:global_Da55}(a) for the particular case $\hat{h}_p = 1$.  Note that this prediction also agrees with an earlier result of \cite{chen_convection_1992} that the two-layer Nusselt number is bounded above by $2/\hat{h}_p$ when the porous layer remains stable. 
%
%

In fact, prediction of $\theta_i$ allows for more accurate prediction of the threshold of convection in the porous medium.  
%
%
Neglecting any temperature variations induced by fluid convection above, we expect that porous convection sets in when $\Ra \Da\, \hp  (1-\theta_i)$ reaches a critical value $\Ra_p^c \simeq 27.1$ \citep{nield_convection_2017}. 
Because $\theta_i \rightarrow 0$ at large Rayleigh numbers while the porous layer remains stable, this condition may be recast as
\begin{equation}
\label{eq:exact_porous_threshold}
\Ra =  \frac{\Ra_p^c}{\Da \,\hp} \left[ 1+ \theta_i  + O(\theta_i^2) \right]
\end{equation}
at the threshold of porous convection. 
To leading order, the threshold is given by $\Ra = \Ra_p^c /(\Da \hp)$, as already noted in \ref{sec:porous_stability}. For the value of $\Da$ used in \ref{sec:overview} where $\hp = 1$, including the first-order correction increases the predicted threshold value of $\Ra$ by about $10\%$.  
%
%

%
%

%

\subsubsection{Asymptotic predictions}

\begin{figure}
\centerline{
\includegraphics[width=0.5\linewidth]{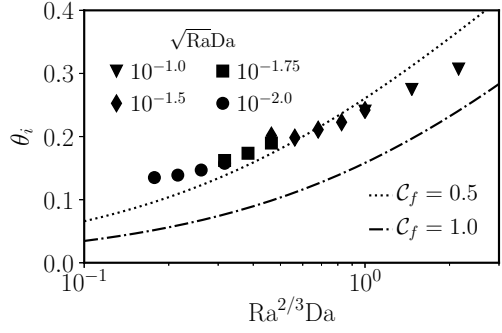}
}
\caption{
Plot of the interface temperature $\theta_i$  found in simulations (symbols) and predicted with the asymptotic model \ref{eq:interface_temperature_asymptotic} as a function of $\Ra^{2/3} \Da$. 
The two lines materialise the two limiting cases of strongly confined porous medium ($\RaDa \rightarrow 0$, $\Cp = 0.85$ and $\Cf = 0.5$) and the weakly confined case ($\RaDa \lesssim 1$, $\Cp = 0.85$ and $\Cf = 1.0$).
Note that  $\Ra^{2/3} \Da$ must remain smaller thant $\sim 10$ to ensure that the constraint \ref{eq:limit_pore_scale} is satisfied. 
}
\label{fig:interface_temperature_asymptotic}
\end{figure}

These relationships for $\theta_i$, and thus for the flux, simplify in the asymptotic regime of very large Rayleigh numbers and extremely low Darcy numbers that is relevant for geophysical applications ($\Ra \Da \gg 1$ while $\RaDa \ll  1$). 
In such a regime, the Rayleigh-Nusselt relations in convecting sub-layers follow the asymptotic scalings discussed at the end of \ref{sec:asymptotic_BL}: 
\begin{equation}
\mathcal{N}_p = 6.9 \times 10^{-3} \hR_p \equiv \alpha_p \hR_p ~~~~\mbox{and}~~~~\mathcal{N}_f \simeq 7 \times 10^{-2} \hR_f^{1/3} \equiv \alpha_f \hR_f^{1/3}.
\end{equation}
In the case of a stable porous medium, flux balance \ref{eq:diffusive_interface_temperature} reduces to 
\begin{equation}
\label{eq:theta_i_stable_asymptotic}
\alpha_f \Ra^{1/3} \theta_i^{4/3} = \frac{1-\theta_i}{\hp }
\end{equation}
with an asymptotic upper bound 
\begin{equation}
\label{eq:stable_upperbound}
\theta_i \sim  \frac{7.3}{\hp \Ra^{1/4}}   
\end{equation}
for $\Ra \gg 1$. In the case where both layers are convecting, \ref{eq:interface_temperature_equation} instead reduces to
\begin{equation}
\label{eq:interface_temperature_asymptotic}
\theta_i^{4/3} = \frac{(2 \mathcal{C}_p)^{2}}{(2 \mathcal{C}_f)^{4/3}} \frac{\alpha_p}{\alpha_f} \, \Ra^{2/3} \Da \,(1-\theta_i)^2,
\end{equation}
which shows that the interface temperature is a function of the grouping $\Ra^{2/3} \Da$ alone in this limit. 
The height of the layers $\hp$ and $\hf$ do not appear in \ref{eq:interface_temperature_asymptotic} because the heat flux is controlled by boundary layers whose widths are asymptotically independent of the depth of the domain \citep{priestley_convection_1954}.

The predictions of \ref{eq:interface_temperature_asymptotic} are shown in \ref{fig:interface_temperature_asymptotic} for the two limiting values of $\Cf$, along with the simulation data for reference. There is rough agreement between data and prediction, although the prediction slightly underestimates the data, presumably owing to the finite values of $\Ra$ and $\Ra_p$ in the simulations.  
The figure demonstrates that $\theta_i$ decreases with decreasing $\Ra^{2/3} \Da$: at constant Rayleigh number, decreasing the Darcy number decreases the efficiency of heat transport in the porous medium and, as a consequence, the porous layer must absorb most of the temperature difference, forcing a decrease in $\theta_i$. In the asymptotic limit of large $\Ra$ but very small $\Da$, such that $\Ra^{2/3} \Da$ remains small, the interface temperature from \ref{eq:stable_upperbound} satisfies
\begin{equation}
\theta_i \sim \sqrt{\Ra \Da^{3/2}}, 
\end{equation}
for $\Ra^{2/3} \Da \ll 1$. Conversely, $\theta_i$ increases with increasing $\Ra^{2/3} \Da$ but evidently cannot increase without bound, which reflects the fact that the grouping $\Ra^{2/3}\Da$ cannot take arbitrarily large values. For $\Ra^{2/3}\Da \gtrsim O(10)$, the flow structures in the porous medium will become smaller than the pore scale, breaking the assumption of Darcy flow there (see \ref{eq:limit_pore_scale}). 

%

%
%

\subsection{Penetrative convection in the porous-diffusive regime}
\label{penetrative_sec}

We end this discussion of fluxes with a  brief consideration of the issue of so-called `penetrative convection' - significant subcritical flow in the porous layer - which has been a contentious subject in some previous studies.
For instance, \cite{poulikakos_high_1986} and more recently \cite{bagchi_natural_2014} reported numerical simulations of porous flows forced by fluid convection despite the porous Rayleigh number being sub-critical, but \cite{chen_convection_1992} found that convection should be confined to the fluid layer only. 
In the simulations detailed in \ref{sec:overview}, we found that while weak flows do exist in the porous-stable case, they do not enhance heat transport compared to diffusion, in opposition to the results of \cite{bagchi_natural_2014} (see for example their figure 3.3).
We argue here that, provided the Darcy number is not too large, this is a general result: it is not possible to induce subcritical flow in the porous layer that has a significant impact on the heat flux through the layer, without violating the limitations of the model outlined in \ref{sec:limits}. 

While the lower boundary condition on the porous layer is uniform, $\theta(z=-\hh_p) = 0$, the temperature at the the interface will, in general, display horizontal variations due to fluid convection above.
Because the temperature cannot be smaller than $0$, the amplitude of these horizontal variations is at most of the order of the interface temperature $\theta_i$.
Any penetrative flow must be driven by these horizontal variations, and so will have an amplitude $w \sim \RaDa \theta_i$. 
%
%
The heat transported on average by the penetrative flow scales with $ w \theta \sim \RaDa \theta_i^2$, and so its contribution to the Nusselt number $\Nu$ is $\sim \Ra \Da \theta_i^2$. 
The contribution of the penetrative flow thus becomes significant, relative to the $O(1)$ diffusive flux through the layer, when $\Ra \Da \theta_i^2 \sim O(1)$. 
But according to the upper bound in \ref{eq:stable_upperbound}, $\Ra \Da \theta_i^2 \lesssim 50 \sqrt{\Ra} \Da$ for $\Ra \gg 1$. 
The assumption that penetrative flows do not transport appreciable heat is therefore accurate for as long as $50 \RaDa \lesssim 1$, or $\Ra_p \lesssim 4 \times 10^{-4} \Da^{-1}$.  Provided $\Da < O(10^{-5})$, this value of the porous Rayleigh number is always larger than the critical value $\Ra_p^c$ for the onset of convection in the porous layer, and there will be no enhanced penetrative convection in the porous layer at subcritical values of $\Ra_p$. 
%
%

Given that in most geophysical systems we expect $\Da \ll 10^{-5}$, we conclude that in general diffusion controls the heat flow through the medium and penetrative convection plays a negligible role. In general, for $\Ra \gg 1$ penetrative convection can only occur if the Darcy number is such that either the constraint $\RaDa \ll 1$ \ref{constraint1} - which enforces that the flow in the porous medium remains confined with negligible inertial effects - or the constraint $\Ra \Da^{3/2} \lesssim 50$ \ref{eq:limit_pore_scale} - which enforces that the flow lengthscales are larger than the pore scale - are violated.

%

\section{Temporal coupling between the layers}
\label{sec:temporal}

In this two-layer set-up, heat is transported through two systems with very different response timescales while carrying the same average heat flux. 
The dynamics in the unconfined fluid layer must, therefore, exhibit variability on both a slow timescale imposed by the porous layer below and on a rapid timescale inherent to turbulent fluid convection.
In turn, as heat is transported to the top of the two-layer cell, fluid convection must mediate and possibly filter the long variations of the porous activity in a manner that remains to be quantified.

\subsection{Heat-flux variations with height}
The contrast between imposed and inherent variability of fluid convection is first illustrated by time series of the horizontally averaged heat flux $J(t,z) = \overline{w\theta} - \Pei \overline{\theta}'$ at different heights, shown in \ref{fig:heat_fluxRa8_Da55}.
The flux in the porous layer experiences long-lived bursts of activity that can amount to up to a 50\% increase of the flux compared to its average value, with a duration that is controlled by the porous turnover timescale $\tau_p \sim (\sqrt{\Ra\Da})^{-1}$. 
In \ref{fig:heat_fluxRa8_Da55}, the signature of these long-time variations can be traced up to the top of the fluid layer, where they are superposed on much faster variations in heat flux associated with the turbulent convective dynamics, which evolve on the $O(1)$ free-fall timescale. 
However, comparison between the time series at $z=0$ and $z=1$ in \ref{fig:heat_fluxRa8_Da55} reveals that the typical intensity of the bursts is notably weaker at the top of the fluid layer than at the interface. 
Therefore, fluid convection is not a perfect conveyor of the long-time, imposed variability, which it partially filters out. 
%

%

\subsection{Spectral content of the heat flux at different heights}

\begin{figure}
\centerline{
\includegraphics[width=\linewidth]{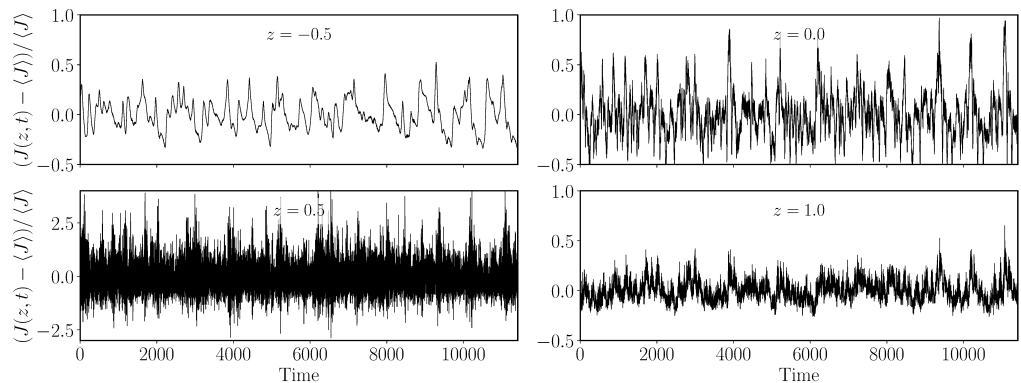}
}
\caption{Time series of the $x$-averaged flux $J (z,t) = \overline{w\theta }-  \Pei \overline{\theta}'$ at different heights in a simulation carried out at $\Ra = 10^8$ and $\Da = 10^{-5.5}$.
The heat flux is normalised by its averaged value over the whole domain and over time. Note that the scale of the $y$ axis is larger at mid-height in the fluid layer ($z = 0.5$).
Time is normalised by the free-fall timescale. The present simulation spans over more than one diffusive timescale since $\taud = 10^4$.
 }
\label{fig:heat_fluxRa8_Da55}
\end{figure}

We use spectral analysis of the heat flux time series to better quantify the inherent and imposed variability of fluid convection and the latter's filtering effect on imposed variability.
\Cref{fig:decay_function}(a) shows the power spectra $\vert \hat{J}(\omega,z)\vert^2$ of the signals displayed in  \ref{fig:heat_fluxRa8_Da55}.
First, the spectral content that is inherent to fluid convection is easily distinguished from the one that is imposed by porous convection.
The variability of the flux in the fluid layer ($z=-0.5$ in \ref{fig:decay_function}(a)) is almost entirely contained in harmonics smaller than $\sim 10^{-2}$.
The energy of lower harmonics in the fluid layer ($z\geq 0$) closely follows the spectral content in the porous layer, which confirms that it is primarily imposed by the porous flow.  
Higher harmonics are therefore controlled by fast fluid convection.
This is particularly well illustrated by the spectra at the fluid boundaries ($z=0$ \& $z=1$) being effectively identical above $\omega = 5 \times 10^{-2}$: we retrieve the top-down symmetry of classical Rayleigh-B\'enard convection with imposed uniform temperature at the boundaries for these larger harmonics. 
The filtering effect of the fluid layer is visible at lower frequencies where the energy decays as $z$ increases.
It is further quantified in \ref{fig:decay_function}(b) where we show the energy at particular frequencies as a function of depth through the fluid layer. 
We find that at low frequency, the energy decays exponentially with $z$ and with a rate that seems to increase with $\omega$. 
The spatial decay is lost for higher harmonics, which are driven directly by fluid convection; instead, the energy is maximised in the bulk of the fluid layer and decays at the boundaries.

\subsection{Spatial decay rate of low-frequency variability}
\begin{figure}
\includegraphics[width=\linewidth]{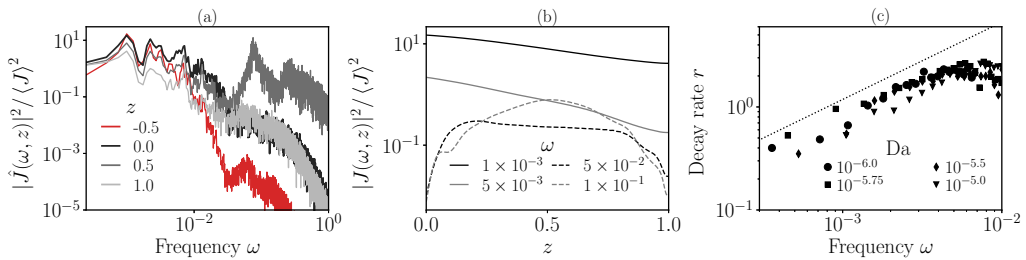}
\caption{
\textbf{a:} Power spectrum $\vert \hat{J}(\omega,z) \vert^2$ of the heat flux timeseries shown in figure \ref{fig:heat_fluxRa8_Da55} and normalised by $\Ja^2$. 
\textbf{b:} Spatial variations of the power spectrum $\vert \hat{J}(\omega,z) \vert^2$ in \textbf{a} with depth through the fluid layer, shown for several frequencies, the lower two being typical of porous convection and the larger two typical of fluid convection. 
\textbf{c:} Vertical decay rate $r$ of the energy of low frequency harmonics of the heat flux, for $\Ra = 10^{8}$ with $\Da$ varied by an order of magnitude from $10^{-6}$ to $10^{-5}$ as indicated in the legend. 
The dotted line gives the power law $\omega^{0.75}$ for reference.  
}
\label{fig:decay_function}
\end{figure}

We attempt to quantify the filtering effect of fluid convection on the long-time variations by systematically measuring the spatial decay rate of the low frequency harmonics. 
By linear fitting of $\ln (\vert \hat{J} (\omega,z)\vert^2 )$ with $z \in [0,0.6]$, we extract the decay rate $r$ for all frequencies below $\omega =  10^{-2}$, above which temporal variations become partially imposed by fluid rather porous convection. 
The result of this process is shown in \ref{fig:decay_function}c.  Despite some spread in the extracted values, they all appear to follow the same trend: 
below $\omega \simeq 5 \times 10^{-3}$, the energy decay rate roughly increases like $\omega^{0.75}$, and above this value the decay rate saturates, presumably because the harmonics are increasingly driven by fluid convection. That is, sufficiently slow variations imposed by the porous layer decay very slowly through the fluid layer, but more rapid variations decay faster. 
The exponential decay of the flux harmonics with $z$ is reminiscent of the problem of diffusion in a solid submitted to an oscillating temperature boundary condition. 
In that classical problem the decay rate has a diffusive scaling, $r\sim \omega^{1/2}$ for oscillation frequency $\omega$.
Here, our results instead suggest that fluid convection acts as a sub-diffusive process on the low frequency flux variations imposed by the porous medium. 
A quantitative explanation for such behaviour remains elusive. It does, however, at least seem reasonable that the decay rate should increase with $\omega$. As $\omega \to 0$ the decay rate must vanish, since the average heat flux through the system is conserved with height, but higher frequency variations in the porous layer will lead to localised bursts of plumes that are quickly mixed into the large-scale circulation of the fluid convection; the extra flux momentarily increase the temperature of the fluid but is not transmitted up to the top of the layer. 
%
%
%
Additional work, possibly with more idealised models of fluid convection submitted to flux variations, is required to fully understand the phenomenology that we have outlined here. 
Although our results on the filtering effect remain preliminary, we can still predict the typical porous turnover timescale $\tau_p$ needed to ensure that the decay of the variability in the fluid layer is negligible. 
The power spectrum of the flux at $z=0$ and $z=\hf$ are in a ratio $\sim \exp(-r(\omega) \hf)$.
The cut-off frequency of such a low-pass filter, reached when $\exp(-r \hf) \sim 1/2$ is roughly located at $\omega = \omega_c = 10^{-3}$ in the particular case $\hf = 1$ according to \ref{fig:decay_function}(c).
Therefore, using $r \sim \omega^{3/4}$, we predict in general that when $1/\tau_p = \RaDa$ is smaller than $\sim \hf^{-4/3} \omega_c$, the temporal variability imposed by porous convection is sufficiently slow that it will be entirely transmitted across the fluid layer. 
%
%

%

\section{Conclusion}
\label{sec:conclusion}

In this article, we have explored heat transport in a Rayleigh--B\'enard cell comprised of a fluid-saturated porous bed overlain by an unconfined fluid layer, using numerical simulations and theoretical modelling. 
The focus of the work has been on the geologically relevant limit of large Rayleigh number, $\Ra$, and small Darcy number, $\Da$, such that heat is transported through the system by vigorous convection but the flow within the porous medium remains inertia-free and well described by Darcy's law. To the best of our knowledge, this is the first study of porous-fluid two-layer systems in this limit. 

Having identified suitable limits on the parameters, we demonstrated that the dynamics and heat flux through the two-layer system is strongly dependent on whether the flow in the porous layer is unstable to convection, which we demonstrated should occur if $\Ra \gtrsim 27/(\hp\Da)$. By suitably rescaling, we showed that flux laws from individual convecting fluid or porous layers could be used to predict both the flux through the two-layer system and the mean temperature at the interface between the two layers. In the asymptotic limit of large $\Ra$, we find that while flow in the porous layer remains stable ($\Ra\Da\hh_p \lesssim 27$) the interface temperature $\theta_i$ satisfies $\theta_i \sim \Ra^{-1/4}$, whereas if the porous layer becomes unstable to convection, $\theta_i \sim \sqrt{\Ra \Da^{3/2}}$, provided $\Da$ remains sufficiently small that $\Ra \Da^{3/2} \ll 1$. 

%
%
%
%
%

We briefly investigated the role played by ``penetrative convection'' \citep{bagchi_natural_2014}, \ie subcritical flows in the porous medium driven by fluid convection above. 
If the fluid above is convecting, then some weak flow will always be driven in the porous layer by horizontal temperature variations imposed at the interface; but we show that they are always too weak to contribute significantly to heat transport, unless some of the model assumptions about flow in the porous medium are violated.
\textcolor{RED}{
Interestingly, the laws governing or limiting heat transport and the interface temperature in porous-fluid convection have been derived from the behaviour of each layer considered separately.
They do not rely on the details of the flow at the interface, in particular on the pore-scale boundary layer at the transition between the porous and the pure fluid flows.
Therefore, the laws that we have identified here should only marginally depend on the choice of the formulation of the two-layer convection problem (see the discussion in \ref{sec:DarcyBrinkmanmodel}). 
}

Lastly, we also briefly explored the manner in which rapid fluid convection mediates the long-time variations of activity in the porous layer.
The amplitude of low-frequency temporal variations of flux imposed by porous convection decay exponentially through the unconfined fluid layer in a way that is reminiscent of a diffusive process. 
As a result, fluid convection acts as a low-pass filter on the bursts of activity in the porous layer.
We predict that for $\RaDa < 10^{-3} \hf^{-4/3}$, the decay of low-frequency variations in the fluid layer are negligible and the variability of porous convection is entirely transmitted to the top of the two-layer system. 

Before ending, we return to consider the question of how important it is to resolve both porous and fluid layers when studying these coupled systems in astrophysical or geophysical settings, rather than using parameterised boundary conditions. Our results in the limit of strong convection suggest that, while the details of convection in each layer are important for controlling the interface temperature and heat flux through the system, there is very little coupling in the dynamical structure of the flow between each layer.  As may be noticed in \ref{fig:snapshots_Da1e_5_5}, for example, fluid convection remains organised in large-scale rolls even if it is forced by several hot plumes from the porous medium below. In other words, a localised `hot spot' at the interface associated with, say, a strong plume in the porous medium, does not, in general, lead to any associated hot spot at the surface of the fluid layer. 

This observation helps to justify the approach of various studies which neglect the dynamics of flow in the porous medium altogether (e.g. \cite{soderlund_ocean_2019} and \cite{amit_cooling_2020} in the context of convection in icy moons). It is also an interesting point in the context of Enceladus, which is well known for sustaining a strong heat-flux anomaly at its South Pole that is believed to be due to hydrothermal circulation driven by tidal heating in its rocky porous core \citep{spencer_cassini_2006,choblet_powering_2017,spencer_plume_2018,le_reun_internally_2020}. While models predict that tidal heating does drive a hot plume in the porous core at the poles \citep{choblet_powering_2017}, our results suggest that this hot spot is unlikely to be transmitted through the overlying ocean to the surface without invoking other ingredients such as rotation \citep{soderlund_ocean-driven_2014,soderlund_ocean_2019,amit_cooling_2020} or topography caused by melting at the base of the ice shell \citep{favier_rayleighbenard_2019}.

\backsection[Acknowledgements]{The authors are grateful to Eric Hester for his help in implementing the code \textsc{Dedalus}. }

\backsection[Funding]{This work was performed using resources provided by the Cambridge Service for Data Driven Discovery (CSD3) operated by the University of Cambridge Research Computing Service (www.csd3.cam.ac.uk), provided by Dell EMC and Intel using Tier-2 funding from the Engineering and Physical Sciences Research Council (capital grant EP/P020259/1), and DiRAC funding from the Science and Technology Facilities Council (www.dirac.ac.uk). 
TLR is supported by the Royal Society through a Newton International Fellowship (grant nr. NIF\textbackslash R1\textbackslash 192181). 
}

\backsection[Declaration of interests]{The authors report no conflict of interest.}

\end{document}